\documentclass[conference]{IEEEtran}
\IEEEoverridecommandlockouts
\usepackage{cite}
\usepackage{amsmath,amssymb,amsfonts}
\usepackage{algorithmic}
\usepackage{graphicx}
\usepackage{textcomp}
\usepackage{xcolor}
\usepackage{soul}

\def\BibTeX{{\rm B\kern-.05em{\sc i\kern-.025em b}\kern-.08em
    T\kern-.1667em\lower.7ex\hbox{E}\kern-.125emX}}

\title{A multi-modal approach for identifying schizophrenia using cross-modal attention\\
\thanks{This work was supported by the National Science Foundation grant numbered 2124270.}
}

\makeatletter
\newcommand{\linebreakand}{%
  \end{@IEEEauthorhalign}
  \hfill\mbox{}\par
  \mbox{}\hfill\begin{@IEEEauthorhalign}
}
\makeatother
\author{
\IEEEauthorblockN{Gowtham Premananth}
\IEEEauthorblockA{\textit{Department of Electrical and Computer Engineering} \\
\textit{University of Maryland, College Park}\\
Maryland, USA \\
gowtham8@umd.edu}
\and
\IEEEauthorblockN{Yashish M.Siriwardena}
\IEEEauthorblockA{\textit{Department of Electrical and Computer Engineering} \\
\textit{University of Maryland, College Park}\\
Maryland, USA \\
yashish@umd.edu}
\linebreakand
\IEEEauthorblockN{Philip Resnik}
\IEEEauthorblockA{\textit{Institute for Advanced Computer Studies} \\
\textit{University of Maryland, College Park}\\
Maryland, USA \\
resnik@umd.edu}
\and
\IEEEauthorblockN{Carol Espy-Wilson}
\IEEEauthorblockA{\textit{Department of Electrical and Computer Engineering} \\
\textit{University of Maryland, College Park}\\
Maryland, USA \\
espy@umd.edu}
}
\begin{document}

\maketitle

\begin{abstract}
This study focuses on how different modalities of human communication can be used to distinguish between healthy controls and subjects with schizophrenia who exhibit strong positive symptoms. We developed a multi-modal schizophrenia classification system using audio, video, and text. Facial action units and vocal tract variables were extracted as low-level features from video and audio respectively, which were then used to compute high-level coordination features that served as the inputs from the audio and video modalities. Context-independent text embeddings extracted from transcriptions of speech were used as the input for the text modality. The multi-modal system is developed by fusing a segment-to-session-level classifier for video and audio modalities with a text model based on a Hierarchical Attention Network (HAN), with cross-modal attention. The proposed multi-modal system outperforms the previous state-of-the-art multi-modal system by 8.53\% in the weighted average F1 score.
\end{abstract}


\vspace*{-3pt}
\section{Introduction}
\vspace*{-3pt}

Schizophrenia is a complex and chronic neuropsychiatric disorder characterized by symptoms with varying severity that result in changes in behavior and impairments in the way reality is perceived by those affected \cite{hansen2023automated}. It affects around 24 million people worldwide \cite{institute2021global}. Schizophrenia is categorized into positive, negative, and mixed schizophrenia, based on characterizing symptoms following Andreasen and Olsen \cite{10.1001/archpsyc.1982.04290070025006}. Positive symptoms of schizophrenia are characterized by disorganization in the expression of thoughts, delusions, hallucinations, and persistent bizarre behavior, while negative symptoms are categorized by attentional impairment, poverty of speech, lack of motivation, blunted affect, and reduced ability to experience pleasure; mixed schizophrenia prominently displays either positive or negative symptoms, or neither is displayed in a prominent manner. The primary assessment procedures in place for schizophrenia are different scales used to measure the positive and negative symptoms of schizophrenia. These scales require human judgment and their quality and general utility vary, suggesting that automated and objective methods to identify symptoms of schizophrenia could provide significant value to clinicians.

Natural language processing has demonstrated to provide insights into a person's mental state and cognitive function \cite{FOLTZ2022}, leading researchers to investigate biomarkers using language processing and speech analysis to detect symptoms exhibited in various psychiatric disorders, including schizophrenia \cite{CORCORAN2020770}. Follow-up studies have shown that schizophrenia can be identified with better accuracy using objective multi-modal approaches \cite{siriwardena2021multimodal, HPERS2021InvertedVT}, involving audio and video modalities, rather than using uni-modal approaches. However, most of the previous work is limited to two modalities and still has room for performance improvements.

In this work, we take advantage of attention mechanisms in deep neural networks to achieve such improvements. Attention mechanisms \cite{vaswani2017attention,luong2015effective} capture relationships between hidden states to characterize which aspects of a state's representation contribute more toward the final prediction. This can be useful not only in improving prediction performance, but also in helping to explain what aspects of the input may have been most relevant \cite{yang2016hierarchical}. Attention mechanisms have been successfully utilized in improving the performance of multi-modal approaches for detecting depression \cite{saggu2022depressnet}, but to the best of our knowledge, they have not been explored in detecting schizophrenia.

Previous studies have shown promising results in identifying the severity of mental health disorders like major depressive disorder and schizophrenia using the correlation structure of the movements of various articulators \cite{siriwardena2021multimodal,Espy-Wilson2019}. Siriwardena et al. \cite{siriwardena2021multimodal} specifically looked into understanding how positive symptoms of schizophrenia affect the articulatory coordination in speech. These findings are the impetus for the current study, in which we introduce a multi-modal approach by fusing three modalities --- audio, video, and text --- using self and cross-modal attention mechanisms, to detect subjects who display strong positive symptoms in schizophrenia.

\begin{figure*}[th!]
    \centering
    \includegraphics[width=\textwidth, height= 60mm]{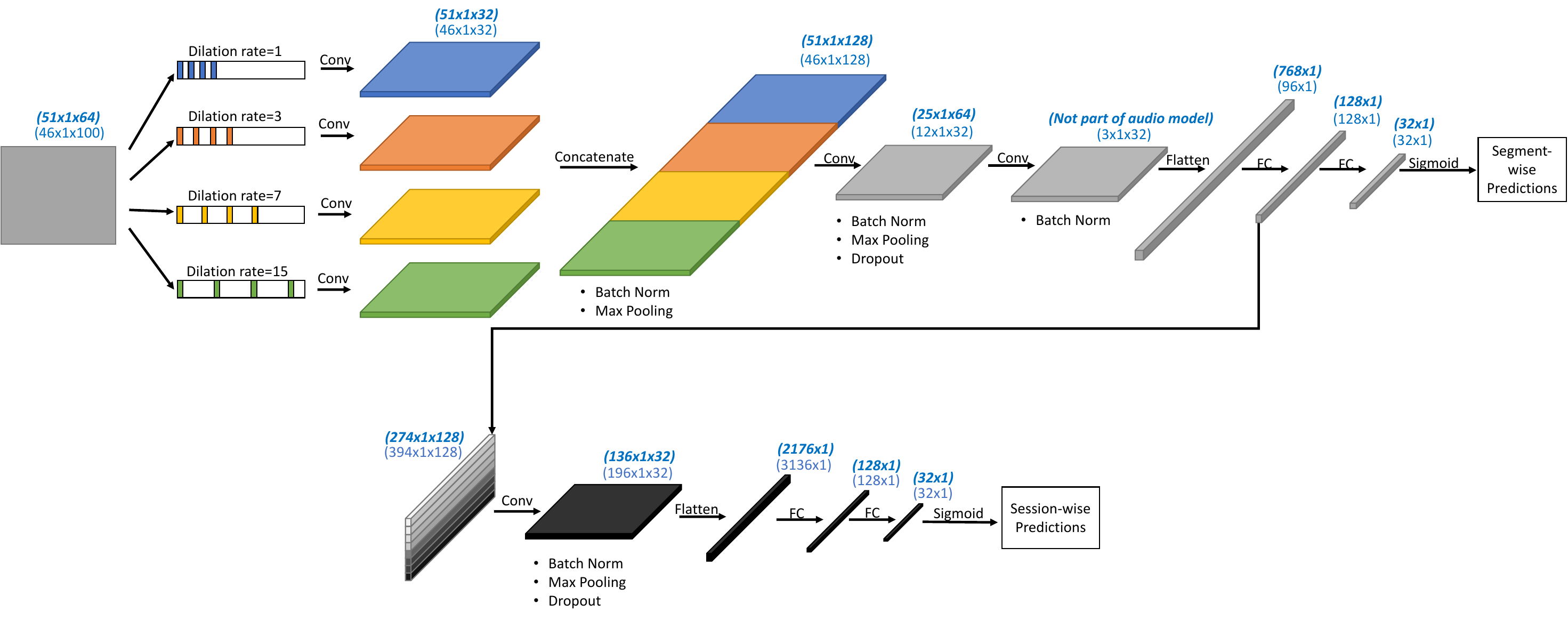}
    \caption{Segment-to-session-level video classification model (\textcolor{blue}{\textbf{\textit{audio model dimensions are denoted on top}}, video model dimensions are denoted below the audio model dimensions)}}
    \label{fig:video}
    \vspace*{-5pt}
\end{figure*}

\vspace*{-3pt}
\section{Database}
\vspace*{-3pt}

Our database was collected as part of an IRB-approved study involving subjects with schizophrenia, depression, and healthy controls, conducted at the University of Maryland School of Medicine in collaboration with the University of Maryland, College Park \cite{KELLY2020113496}. The database consists of video and audio recordings of subjects providing free responses in an interview format. As this particular study focuses just on subjects showing positive schizophrenia symptoms, a subset of suitable subjects (7 schizophrenia subjects, and 11 healthy controls) based on the clinician assessments provided with similar cumulative duration of speech was selected. The selected subset of the database contains a total of 19.43 hours of 50 unique interview sessions belonging to the 18 selected subjects.

\vspace*{-3pt}
\section{Feature extraction}
\vspace*{-3pt}

The audio recordings in the database include the speech of both the interviewer and the subjects and are transcribed and diarized manually by a third-party transcription service. The speech corresponding to the subject is extracted (using the time stamps in the transcripts) and are segmented into small chunks with a minimum duration of 5~seconds as in \cite{siriwardena2021multimodal}.


As audio-based features, vocal tract variables (TVs) were extracted from the acoustic signals of the segmented audio recordings using an acoustic-to-articulatory speech-inversion system \cite{siriwardena2022secret}. A total of 6 TVs were extracted. In addition, two other glottal parameters, aperiodicity, and periodicity were also extracted using an Aperiodicity, Periodicity, and Pitch detector \cite{deshmukh}. This voicing information has been shown to improve the accuracy of the speech inversion system that estimates the 6 TVs \cite{siriwardena2022secret} and the accuracy of mental health classification systems\cite{seneviratne2020extended}. Since previous studies in detecting mental health disorders (depression and schizophrenia) \cite{siriwardena2021multimodal, 10.1145/3347320.3357688} have shown that the Mel-Frequency Cepstral Coefficients (MFCCs) and TVs outperform features like extended Geneva Minimalistic Acoustic Parameter Set (eGeMAPS) \cite{7160715} and DEEP SPECTRUM features \cite{amiriparian17_interspeech} in audio-based prediction tasks, those features were not used. However, baseline models were also trained using self-supervised audio features like Wav2Vec \cite{10.5555/3495724.3496768} and HuBERT\cite{DBLP:journals/corr/abs-2106-07447} for comparison. 

Openface~2.0: Facial Behaviour Analysis toolkit \cite{baltrusaitis2018openface} was used to extract 10 Facial Action Units (FAUs) that capture the coordination of lip and near-lip movements when the subjects are speaking \cite{prince2015facial} from the video recordings. The extracted FAUs were used for the classification models.

From the TVs extracted from audio and the FAUs extracted from video, a high-level correlation structure was computed based on the work in Huang et al. \cite{huang2020exploiting}, which calculates correlations starting from 0 to a delay of 'D' frames. The delayed autocorrelation and cross-correlations across TVs in the segments are stacked together to create the Full Vocal Tract Coordination (FVTC) correlation structure. A grid search from (45,50,55) was done to pick the 'D' parameter used to create the correlation structures and the best unimodal performance was achieved with D=50 for TVs and D=45 for the FAUs. 

From the transcriptions of the sessions in the dataset, the subjects' speech was separated based on the speaker IDs provided. Then for text pre-processing, the punctuation and stop words were first removed and the text was tokenized using the NLTK (English) tokenizer. Finally, context-independent 100-dimensional GloVe word embeddings \cite{pennington2014glove} were extracted to be used as input for the text-based classification model.

\vspace*{-5pt}
\section{Methodology}
\vspace*{-3pt}

\subsection{Uni-modal model architecture} 
\vspace{-5pt}
 
Using the extracted coordination features as input, separate segment-to-session-level classification neural network (STS-CNN) models as shown in Figure \ref{fig:video} were designed and trained to classify the subjects based on uni-modal video and audio data. For segment-level classification, we created a Dilated Convolution Neural Network (DCNN) architecture where the delay
scales were chosen while considering each scale to be roughly twice as big as the previous
scale, and to sample as many as possible different points along the auto and cross-correlation matrix following the work of Huang et al. \cite{huang2020exploiting}.  As the goal is to perform session-level classification, the output of the first fully connected layer in the segment-level classifier is taken across all segments of a session, stacked together, and then sent through a convolution neural network which produces a session-level label. 

As the language model for transcribed text data, a bidirectional LSTM-based hierarchical attention network (HAN) architecture was utilized. The HAN model applies the attention mechanism at both word and sentence levels, thus taking into consideration the hierarchical structure of the text in the session.

\vspace{-5pt}
\subsection{Attention Mechanism}
\vspace{-5pt}

Our multi-model architecture utilized multiple attention mechanisms (self-attention and cross-modal attention) designed based on the dot product attention \cite{luong2015effective} to improve multimodal prediction. Dot product attention is calculated as shown in Eq. \ref{eq:attention} where $Q$ denotes the Query matrix, $K$ denotes the key matrix with a dimension of $d_K$, and $V$ is the Value matrix.

\vspace*{-3pt}
\begin{equation}
 Attention(Q,K,V) = softmax(\frac{QK^T}{\sqrt{d_K}})V
\label{eq:attention}
\end{equation}
\vspace*{-3pt}

In self-attention, all of the Query, Key, and Value matrices are the same as the self-attention mechanism tries to find the interactions within the input itself to find out which parts of the input contribute more towards the prediction. In contrast, in cross-modal attention, the Query and Key will be different but the Key and Value will be the same matrices. Cross-modal attention identifies how different modalities affect the prediction of the classes using other modalities when they are fused together.

\vspace{-5pt}
\subsection{Multi-modal model architecture} \label{arch}
\vspace{-5pt}

\begin{figure*}[th!]
    \centering
    \includegraphics[width=\textwidth, height= 44mm]{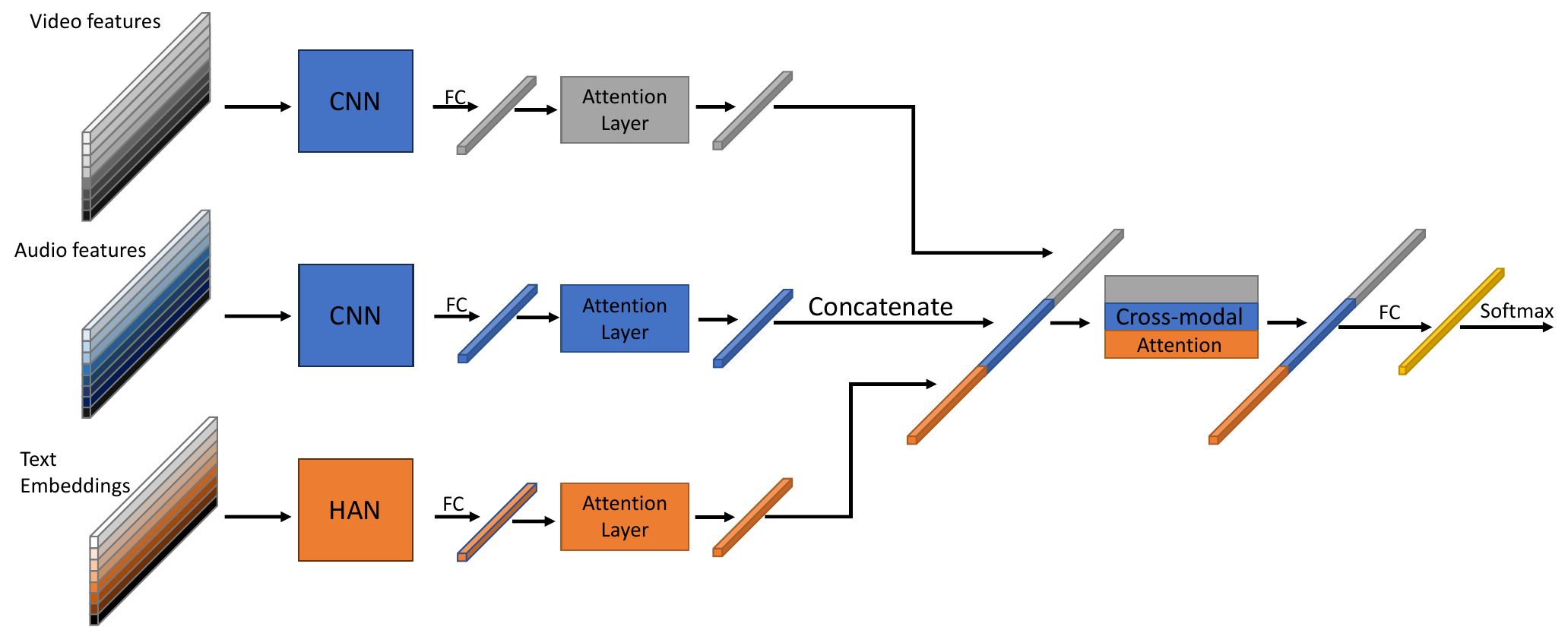}
    \caption{Multimodal classification model}
    \label{fig:multi}
    \vspace*{-3pt}
\end{figure*}

We investigated various multi-modal models by fusing pairs of all the available modalities to check how the better-performing modalities improve the performance of the classifier in the multi-modal setup. We created a multi-modal model (see Fig. \ref{fig:multi}) which gets the stacked session-level audio and video features, along with the tokenized transcribed text of the session as inputs. The model comprises CNN branches to process the audio and video features while a HAN branch was used to process the textual features before fusing. Then all three modalities are sent through self-attention layers, concatenated together, and sent through a cross-modal attention layer which is finally passed through fully connected layers to produce the session-level multi-modal classification output. An ablation study was also carried out for the multi-modal architecture to validate the contribution of different attention mechanisms used in the model.

\vspace*{-3pt}
\section{Experiments and Results}\label{results}
\vspace*{-3pt}

We did a grid search to fine-tune hyperparameters for all the uni-modal and multi-modal classification models. For the video-based and audio-based uni-modal classification models, different segmentation durations (20,30,40 seconds) were experimented with. Out of them, 20-second segmentation and 40-second segmentation with 5-second overlaps showed the best performance for the video and audio models, respectively. 
In \cite{siriwardena2021multimodal}, which was conducted using the same dataset, a leave-one-out cross-validation was used where ideally more data is utilized to train the models, and a comparatively very small portion of the dataset is used for testing. To overcome this data disparity between training and testing of the models, all the models in this study were trained and evaluated based on a 5-fold cross-validation where the models were trained with the input of 3 folds of subjects while being validated on the $4^{th}$ fold and then being tested on the $5^{th}$ fold. For the 5-fold cross-validation, the folds were manually designed to include subjects from both the classes belonging to each of the folds with similar duration of data available in all the folds. Accuracy and F1 scores for each class are used as the evaluation metrics to quantify the performance of these models and they are calculated across all the 5 folds. All the models were trained for 300 epochs with early stopping based on monitoring the validation loss change with patience of 20 epochs. The best performing uni-modal models compared to the baselines are shown in Table \ref{tab:singlemodalities}.

\vspace*{-5pt}
\begin{table}[h]
\centering
\caption{Performance comparison of uni-modal models.}
\footnotesize
  \label{tab:singlemodalities}
    \vspace*{-5pt}
  \begin{tabular}{|l|l|l|c|c|}
    \hline
    Modality& Features& Model&Accuracy&\footnotesize F1(S)/F1(H)\\
    \hline
    Video & \footnotesize FAU-ACF&\footnotesize FVTC-CNN \cite{siriwardena2021multimodal} & 84.55 &0.74/0.88   \\
    \hline
    \textbf{Video}& \footnotesize \textbf{FAU-ACF}&\footnotesize \textbf{STS-CNN}& \textbf{88.68} &\textbf{0.85/0.86}\\
    \hline
    Audio&\footnotesize Wav2Vec2&\footnotesize Wav2Vec2 \cite{10.5555/3495724.3496768}  &65.10 &0.50/0.69 \\
    \hline
    Audio&\footnotesize Hubert&\footnotesize Hubert \cite{DBLP:journals/corr/abs-2106-07447}  &67.96  &0.53/0.70 \\
    \hline 
    Audio&\footnotesize MFCC&\footnotesize FVTC-CNN \cite{siriwardena2021multimodal}  &72.22  &0.55/0.80 \\
    \hline    
    Audio&\footnotesize FVTC&\footnotesize FVTC-CNN \cite{siriwardena2021multimodal}  & 72.22 &0.62/0.78  \\
    \hline
    \textbf{Audio}&\footnotesize \textbf{FVTC}&\footnotesize \textbf{STS-CNN}& \textbf{85.05 }&\textbf{0.77/0.87}\\
    \hline
    \textbf{Text}&\footnotesize \textbf{GLOVE}&\footnotesize \textbf{HAN} & \textbf{79.09}&\textbf{0.71/0.79}\\
      \hline
\end{tabular}
\end{table}

\vspace*{-10pt}
\begin{table}[h]
\centering
  \caption{Performance comparison of Multi-modal models}
  \label{tab:comp_multimodalities}
    \vspace*{-5pt}
  \begin{tabular}{|l|c|c|c|}
  \hline
  Model&Accuracy&F1(S)&F1(H)\\
    \hline
    Video+Audio \cite{siriwardena2021multimodal} & 85.23 & 0.80  &0.83 \\
    \hline
    Video+Audio [proposed] &\textbf{90.68\textbf}  &\textbf{0.92} &\textbf{0.87} \\
    \hline
    Text+Audio [proposed] &87.05  &0.82  &0.84 \\
    \hline
    Video+Text [proposed] &86.68 &0.86 &0.83 \\
    \hline
    \textbf{Multi-modal [proposed]} &\textbf{92.50} &\textbf{0.95} &\textbf{0.88}\\
\hline
\end{tabular}
\end{table} 
\vspace*{-5pt}

One of the key contributions of this work is the introduction of attention mechanisms to enhance the final model prediction. To that end, experiments were carried out by adding different attention mechanisms to the multi-modal models to see whether the added attention mechanisms could necessarily enhance the final predictions compared to the uni-modal systems. As shown in Table \ref{tab:comp_multimodalities}, multiple multi-modal models were designed and tested by fusing all combinations of uni-modal systems. The table also compares the performances of the multi-modal models proposed in this study with the study \cite{siriwardena2021multimodal} conducted using the same dataset.

\vspace{-5pt}
 \subsection{Ablation study} \label{err}
\vspace{-5pt}

We investigated the contribution of the attention mechanisms used in the multi-modal(MM) architecture, by training and testing models with and without the attention mechanisms to validate their importance. The results of the ablation study are tabulated in table \ref{tab:ablation}

\vspace*{-5pt}
\begin{table}[h!]
\centering
  \caption{Ablation study results}
  \label{tab:ablation}
  \vspace*{-5pt}
  \begin{tabular}{|l|c|c|c|}
  \hline
  Model&Accuracy&F1(S)&F1(H)\\
    \hline
    \textbf{MM [proposed]} & \textbf{92.50} & \textbf{0.95}  &\textbf{0.88} \\
    \hline
    MM without self attention &88.86  &0.85&0.93 \\
    \hline
    MM without cross-modal attention &85.23 &0.77  &0.90 \\
\hline
\end{tabular}
\end{table} 
\vspace*{-8pt}

 \subsection{Error analysis} \label{err}
\vspace*{-5pt}

We investigated the misclassifications made by the uni-modal models to find out what was causing those models to misclassify certain sessions. Here we present the results specifically based on the text modality where more misclassifications were done when compared to other uni-modal models. To understand these misclassifications, we calculated coherence scores based on the Latent Dirichlet Allocation (LDA) \cite{blei2003latent,syed2017full} model for all of the text transcripts. In this analysis, it was noted that all the healthy controls' sessions that were misclassified as schizophrenia subjects had lower coherence scores than the average coherence score corresponding to healthy controls. Table \ref{tab:coherence} provides a summary of these coherence scores.

\vspace*{-5pt}
\begin{table}[h]
\centering
  \caption{Coherence scores of the text transcripts}
  \label{tab:coherence}
  \vspace*{-5pt}
  \begin{tabular}{|l|c|}
\hline
    Sessions&Coherence Score\\
    \hline
    All Healthy Controls (Average)&0.5734\\ 
    \hline
    All Schizophrenia subjects (Average)&0.5580\\
    \hline
    Misclassified Healthy controls (range)&0.4706 - 0.5678\\
  \hline
\end{tabular}
\end{table}

All the sessions misclassified by the best-performing audio model were also misclassified by the video model which suggests that both the modalities lacked the necessary information to make an accurate decision. Future work will be done to understand in detail why certain sessions were misclassified by both these modalities so that further improvements can be made in the respective models. 

\section{Discussion}
\vspace*{-3pt}

In this work, we have experimented with three key modalities in speech communication, to distinguish subjects who show strong positive symptoms of schizophrenia from healthy controls. We developed individual uni-modal systems for all three modalities and introduced a fusion strategy involving self and cross-modal attention to leverage cross-modality information to improve schizophrenia detection. This study is inspired by the work in \cite{siriwardena2021multimodal} where only audio and video data from the same dataset was used. As shown in Table \ref{tab:singlemodalities}, the segment-to-session-level classifier model we deployed for the audio and video uni-modal systems outperformed the models deployed in \cite{siriwardena2021multimodal} for the same respective modalities. The multi-modal system fusing audio and video with self and cross-modal attention also outperformed the best performing multi-modal system reported in \cite{siriwardena2021multimodal}.  

A key observation with the current uni-modal systems is that the performance of both the video and audio models has improved while the audio model has improved drastically when compared to the previous work in \cite{siriwardena2021multimodal}. The main reasons for that may be the use of an improved speech inversion system\cite{siriwardena2022secret} to extract the vocal tract variables and the changes adopted in pre-processing with the use of speech segments with overlaps. The use of speech segments with overlaps also reasonably increased the number of samples used to train both audio and video modalities. 

Another significant contribution of the current work comes with the incorporation of a text modality to the mix of audio and video modalities, which to the best of our knowledge has not been experimented at all together in schizophrenia research. The text model produced the lowest performance out of the three individual modalities. This may be because the data was collected in an interview setting and then diarized to get the subject's speech to produce text data that had numerous short utterances. Previous text-based studies \cite{bedi2015automated} have shown, that semantic coherence is one of the key properties of text in differentiating schizophrenia subjects from healthy subjects. But in our case, the subjects' responses are frequently short, which may have made semantic coherence, or the lack of it, more difficult for the model to detect. This may be why the text model produced the lowest performance out of the three individual modalities. Still, the significant contribution of the current work comes with the incorporation of a text modality to the mix of audio and video modalities, which to the best of our knowledge has not been experimented with altogether in schizophrenia research. 

Our key results in Table \ref{tab:comp_multimodalities} show that the tri-modal model with the combination of both self and cross-modal attention mechanisms between all the modalities has outperformed all the other models. These results indicate that the multi-modal systems can compensate for the uni-modal systems' errors and that multi-modal systems with self and cross-modal attention mechanisms can sufficiently mitigate the misclassifications done by the multi-modal models. The results exhibited in Table \ref{tab:ablation} show that the simple fusion of the three modalities is not sufficient and the use of both attention mechanisms is vital for improved performance.

\vspace*{-3pt}
\section{Conclusion and Future Work}
\vspace*{-3pt}

In conclusion, this paper shows that in distinguishing subjects with strong positive schizophrenia symptoms from healthy controls, segment-to-session-level classifiers show better performance in video and audio modalities when compared to previous classification approaches which use ACFs as input features. Furthermore, we obtain even stronger results using a multi-modal approach combining audio, video, and text using self and cross-modal attention. As future work, we plan to expand our investigation to include the full set of schizophrenia subjects, not just those with strong positive symptoms. In addition, we plan to make more use of self-supervised speech features by developing models that can fully utilize the potential of such features to understand if we can further push the boundaries of the multi-modal systems. Furthermore, we are planning to investigate which parts of each modality are given more attention when making the final classification decisions, in order to better understand the performance of our models, and potentially inform the development of explanatory visualizations for clinicians.

\bibliographystyle{IEEEbib}
\bibliography{refs}

\end{document}